\documentclass[pre,twocolumn]{revtex4-1}

\usepackage{graphicx}% Include figure files
\usepackage{dcolumn}% Align table columns on decimal point
\usepackage{bm}% bold math
\usepackage[utf8]{inputenc}
\usepackage[T1]{fontenc}
\usepackage{mathptmx}
\usepackage{bm}
\usepackage{pdfrender}

\usepackage{amsmath}
\usepackage{mathtools}
\usepackage{amssymb}
\usepackage{tikz}

\usepackage{pst-node}
\usepackage{verbatim}   % useful for program listings
\usepackage{color}      % use if color is used in text
\usepackage{subfigure}  % use for side-by-side figures
\usepackage{hyperref}   % use for hypertext links, including those to external documents and URLs

\usepackage{braket} 

\newcommand{\bnabla}{\mbox{\boldmath $\nabla$}}
\newcommand{\ba}{\begin{eqnarray}}
\newcommand{\ea}{\end{eqnarray}}
\newcommand{\be}{\begin{equation}}
\newcommand{\ee}{\end{equation}}

\begin{document}

%==============================================================
\title{Entropy production of active particles formulated for underdamped dynamics} 
%==============================================================

\author{Derek Frydel}
\affiliation{Department of Chemistry, Universidad Técnica Federico Santa María, Campus San Joaquin, 7820275, Santiago, Chile}

\date{\today}

\begin{abstract}
The present work investigates the effect of inertia on the entropy production rate $\Pi$ for all canonical models of active 
particles for different dimensions and the type of confinement.  To calculate $\Pi$, the link between the entropy production 
and dissipation of heat rate is explored resulting in a simple and intuitive expression. 
By analyzing the Kramers equation, alternative formulations 
of $\Pi$ are obtained and the virial theorem for active particles is derived.  Exact results are obtained for particles in 
an unconfined environment and in a harmonic trap.  In both cases, $\Pi$ is independent of temperature.  For the case
of a harmonic trap, $\Pi$ attains a maximal value for $\tau = \omega^{-1}$ where $\tau$ is the persistence time and 
$\omega$ is the natural frequency of an oscillator.  
For active particles in 1D box, or other non-harmonic potentials, thermal fluctuations are found to reduce $\Pi$.  
\end{abstract}

\pacs{
}

\maketitle
%------------------------------------------------

%\section{Introduction}

\section{Introduction}

In this work we study the entropy production rate $\Pi$ 
\cite{Rosen60,Schnakenberg76,Tome06,Leonid10,Tome12,Cates16,Shankar18,Schmidt19,Pruessner20,Razin20,Pruessner21}
within underdamped dynamics for three canonical models of active particles.   Contributions of inertia are
expected to play role in the regime where the persistence time of active motion $\tau$ is comparable to 
or smaller than the inertial relaxation time $\tau_r$.  Even if this regime may be inaccessible to experimental 
systems, the inclusion of inertia is necessary to recover a correct behavior in the limit $\tau \to 0$.  

Our interest to understand the role of inertia is motivated by a recently 
made observation that for the run-and-tumble (RTP) and active Brownian particle (ABP) models the entropy 
production rate formulated within overdamped dynamics is maximal when $\tau=0$ (implying a maximal distance 
from equilibrium) \cite{Frydel22a}.  At the same time, the stationary distribution recovers a Boltzmann functional 
form (indicative of equilibrium).  The present work shows that this inconsistency of conclusions is eliminated if $\Pi$ 
is defined in the underdapmed dynamic regime, where as a result of inertia $\Pi$ vanishes at $\tau=0$.

The reason why $\Pi$ defined in the overdamped regime attains a maximal value in the limit $\tau\to 0$ is as follows.  
The RTP and ABP active motion is determined by two parameters, the constant magnitude of a swimming velocity $v_0$, 
and the persistence time $\tau$.  In the limit $\tau\to 0$, the swimming direction of an active particle changes very fast.  As a 
result, an active motion fails to produce a net displacement in a particle position.  This, effectively, suppresses active 
motion and explains the emergence of a Boltzmann distribution.  The suppression, however, is only apparent and not 
real, since the magnitude of a velocity in the overdamped regime remains constant.  
{At length scales comparable to 
or smaller than $v_0 \tau$, an active motion can still be detected as a kind of erratic vibration, since the magnitude of 
a velocity of active motion is constant.}  
And if the volume element where this "vibration" occurs is sufficiently 
small for an external potential to remain constant, it is as if an active particle 
found itself in unconfined environment.  Unimpeded by any confinement, the entropy production attains a maximal value.  

With inertia taken into account, in the regime $\tau \lesssim \tau_r$ a swimming velocity fails to adjust itself to a rapidly
changing force direction.  As a result, the magnitude of a swimming velocity is reduced, and in the limit $\tau\to 0$ 
it altogether vanishes.  And once a velocity (due to active motion) is suppressed, the entropy production rate goes to zero, 
identifying the limit $\tau\to 0$ as equilibrium.  

In addition to the above motivation, 
the current study is relevant to a growing interest in active 
particle models with inertia \cite{Lowen19,Sandoval20,Lowen22}.  Such models are more representative of 
particles embedded in low-density environment such as gas.  Examples include mesoscopic dust particles 
in plasmas \cite{Morfill09}, granulars on a vibrating plate \cite{Weber13}, or insect flight at water interfaces 
\cite{Kim16}.

To formulate $\Pi$ within underdamped dynamics, we explore the link 
between the entropy production rate and the heat dissipation rate \cite{Sekimoto98,Sasa05,Maes09,Maes21,Landi21}. 
From the Kramers equation for particles in an external potential, we obtain
a number of alternative formulations of $\Pi$.  One by-product
of this analysis is the derivation of the virial theorem for active particles.  Another interesting result is the 
representation of stationary distributions in $v$-space, for active particles in unconfined environment and 
in a harmonic trap, as a convolution between the Maxwell distribution and the distribution at zero temperature.   
This is possible because in both cases the two random processes are independent.  The independence
of the random processes is lost for other types of confinements.  

Furthermore, this work shows that for particles in a harmonic trap the entropy production is found to be maximal 
at $\tau = \omega^{-1}$, where $\omega$ is the natural frequency of a harmonic potential.  For $\tau < \omega^{-1}$ 
the entropy production quickly decreases and then vanishes at $\tau=0$.  This suggests the presence of two
equilibria, in the limits $\tau\to 0$ and $\tau\to\infty$.  The equilibrium in the limit $\tau\to\infty$ represents
a system in equilibrium with quenched disorder \cite{Frydel21a,Frydel22c}.  
This is different from analysis 
based on overdamped dynamics where the entropy production increases monotonically with decreasing $\tau$ 
and is maximal at $\tau=0$.  Therefore, only a single equilibrium at $\tau\to \infty$ is predicted.

The role of inertia in determining entropy production has been investigated in the past.  In \cite{Celani12,Nakayama13}
this was done for a system in the presence of temperature gradients and in \cite{,Chun15}, for a Brownian 
particle in a harmonic trap and in contact with two heat baths.  
To our knowledge, the role of inertia on $\Pi$ for active particle systems 
has been considered for the first time in \cite{Shankar18} for an unconfined environment 
and in two-dimensions, and without considering the RTP model.  More recently, some aspects of 
inertia within the AOUP model were considered in \cite{Lorenzo21}.

\section{Entropy production as a dissipation of heat}
\label{sec:sec0}

The second law of thermodynamics states that a non-reversible process is characterized by the production of entropy.
The process of entropy production is commonly represented as  
\cite{Prigogine55,Groot62,Glansdorff71,Schnakenberg76,Tome12}
\be
\dot S = \Pi - \Phi, 
\label{eq:dS}
\ee
where $\dot S$ is time derivative of the entropy, 
$\Pi$ is the entropy production rate, and $\Phi$ is the entropy flux in and out of the system that occurs in two forms: 
heat transfer $\dot q/T$ and mass flow. Without mass flow, the entropy flux simply is
\be
\Phi = \frac{\dot q}{T},  
\ee
and for the stationary state $\dot S=0$, Eq. (\ref{eq:dS}) leads to 
\be
T\Pi = \dot q,
\label{eq:Pi0}
\ee
establishing a relation between entropy and heat production \cite{Sasa05,Marconi17,Maes03,Maggi17,Landi21}.

\section{Langevin equation analysis}
\label{sec:sec0}

In this work we consider particles whose velocity evolves according to the following (underdamped) Langevin equation
\be
\tau_r \dot {\bf v}   =   -  {\bf v}   +    \pmb{\eta}   +  \mu {\bf f} + \mu {\bf F}_{}, 
\label{eq:Langevin-0}
\ee
where $\tau_r = m\mu$ is the inertial relaxation time, $m$ is the mass of a particle, and $\mu$ is the mobility.  The two 
stochastic processes are thermal fluctuations $\pmb{\eta}$ and the force ${\bf f}$ that is responsible 
for generating active motion.  The two systematic forces are ${\bf F}_{}$, an external and position dependent force, 
and the drag force ${\bf F}_{d}(t) = -{{\bf v}(t)} {\mu}^{-1}$ due to a surrounding fluid.  
Using a definition of power, ${\bf F}_{d}\cdot{\bf v}$, we get the heat dissipation rate $\dot q(t) = {v^2(t)} {\mu}^{-1}$, 
which is proportional to the kinetic energy of a particle.

Thermal fluctuations $\pmb{\eta}$ are represented as a Gaussian white noise which on average imparts a fixed amount 
of energy $E_k^{eq} = \frac{n}{2} k_BT$, where $n$ is the system dimension, $T$ 
is a thermal temperature, and $k_B$ is the Boltzmann constant.  

Dissipation of energy means absorption of heat by an infinite reservoir so that the temperature of a finite system
does not rise.  The energy that comes from the reservoir in form of thermal fluctuations 
is not really dissipated since at another time it is returned to a particle as thermal noise.  Only the part of a kinetic energy 
that comes from ${\bf f}$ is truly dissipated.  
This means that when calculating the average dissipation of heat, the contribution that comes from thermal 
fluctuations needs to be subtracted.  This results in the following formula of the dissipation of heat:  
\be
T\Pi \equiv  \langle \dot q \rangle =   \frac{1}{\mu} \left [ \langle v^2\rangle - \frac{n D}{\tau_r} \right],   ~~~~ \text{RTP and ABP},
\label{eq:w-0}
\ee
where $D = \mu k_BT $ is the diffusion constant and $\tau_r=m\mu$ is the inertial relaxation time.  
The above formulation is the centerpiece of this article.  In subsequent sections, 
this formula, and its alternative manifestations, will be used to calculate $\Pi$ for different scenarios.  
The fact that contributions of $ \pmb{\eta} $ to the kinetic energy remain the same, regardless of the presence
of active motion or any external potential, is a consequence of $\pmb{\eta}$ being independent of ${\bf f}$ and 
${\bf F}$.   This independence will be demonstrated more rigorously in the next section.

It is possible at this stage to propose an alternative formulations of $\langle \dot q \rangle$ by multiplying the 
Langevin equation in Eq. (\ref{eq:Langevin-0}) by ${\bf v}$ and taking average.  This leads to 
\be
\frac{\tau_r}{2} \left\langle  \frac{d v^2}{dt}  \right \rangle   =   -  \langle v^2 \rangle   +    \langle {\bf v} \cdot \pmb{\eta} \rangle  
+  \mu \langle  {\bf v} \cdot {\bf f} \rangle + \mu\langle {\bf v} \cdot {\bf F}_{}\rangle.
\label{eq:Langevin-1}
\ee
The term on the left-hand-side represents flux of the kinetic energy in and out of the system and for a stationary state
it is zero \cite{Kubo66,Felderhof78,Bhattacharjee}.   The second term on the right hand side evaluates 
to $\langle {\bf v} \cdot \pmb{\eta} \rangle = n D/\tau_r$ \cite{Bhattacharjee} (this relation will be derived again later 
from the stationary Kramers equation).  Finally, the last term representing power input due to an external potential 
vanishes, $\langle {\bf v} \cdot {\bf F}_{}\rangle=0$, since the external force on average does not contribute 
to the production of energy in a stationary state.  (This relation will also be derived later from the Kramers equation.)  

Using these results, Eq. (\ref{eq:Langevin-1}) together with Eq. (\ref{eq:w-0}) leads to another definition of the 
dissipation of heat based on the average power input:
\be
T\Pi \equiv  \langle \dot q \rangle  =   \langle  {\bf v} \cdot {\bf f} \rangle. 
\label{eq:g1a}
\ee
Later we derive two other formulations of $\langle \dot q \rangle$ from the Kramers stationary equation.

The interpretation of heat dissipation as a result of an external time dependent force ${\bf f}$ 
poses no problems for the RTP and ABP models, for which this interpretation was originally intended.
In the case of the AOUP model, the force ${\bf f}$ is represented as a colored noise \cite{Sevilla19,Kaarakka11}
\be
\mu {\bf f}   =    \frac{1}{\tau}  \int_{-\infty}^t ds\, e^{-(t-s)/\tau} \pmb\eta_f(s), ~~~~ \text{AOUP},
\label{eq:Langevin-c}
\ee
where $\pmb\eta_f$ is a white noise where the subscript "f" is used to distinguish this noise from the white noise 
of a reservoir.  But note that in the limit $\tau\to 0$, $\mu {\bf f}   \to \pmb\eta_f$, which recovers the standard Langevin 
equation for a passive Brownian motion, $\tau_r \dot {\bf v}   =   -  {\bf v}   +  \pmb{\eta}   + \pmb\eta_f$, however, 
with two sources of white noise and, as a result, an enhanced thermal temperature $T\to T + T_f$.  

This raises a question:  should we regard $\pmb\eta_f$ as an external time dependent noise, or do we regard it as 
a thermal fluctuation, in which case we must postulate, and justify, the existence of a second reservoir.   To treat 
all models uniformly, in this work we always consider ${\bf f}$ as a force with an external source.  The ambiguity that arises 
for the AOUP model does not affect a number of different relations that are later obtained.   The main motivation of 
this work is to understand inertia effects in the RTP and ABP models, for which there is no ambiguity of interpretation, 
on the entropy production rate.

\section{Unconfined environment}
\label{sec:sec1}

%\textcolor{red}{
To obtain an expression for $\Pi$ as defined in Eq. (\ref{eq:w-0}), we need to calculate the variance $\langle v^2\rangle$, 
which can be obtained by solving the Langevin equation:  
\ba
\langle v^2\rangle   &=&  \frac{1}{\tau_r^2} \int_{-\infty}^{t} ds\int_{-\infty}^{t} ds' \, e^{-(t-s)/\tau_r} e^{-(t-s')/\tau_r} \nonumber\\ 
&\times& \bigg[\langle \pmb\eta(s) \cdot \pmb\eta(s') \rangle + \mu^2 \langle {\bf f}(s)\cdot {\bf f}(s')\rangle   + \mu^2 \langle {\bf F}(s)\cdot {\bf F}(s')\rangle  \nonumber\\
&+& 2 \mu \langle {\bf f}(s) \cdot \pmb \eta(s') \rangle + 2 \mu \langle {\bf F}(s) \cdot \pmb \eta(s') \rangle + 2 \mu^2 \langle {\bf F}(s) \cdot {\bf f}(s') \rangle \bigg].  \nonumber\\
\label{eq:v2t}
\ea
Because thermal fluctuations are independent of all the forces, $\langle {\bf f}(t) \cdot \pmb \eta(t') \rangle = \langle {\bf F}(t) \cdot \pmb \eta(t') \rangle = 0$, 
the contributions of thermal fluctuations to $\langle v^2 \rangle$ are always the same and is given by
\be
\langle v^2\rangle_{eq}  = \frac{1}{\tau_r^2} \int_{-\infty}^{t} ds\int_{-\infty}^{t} ds' \, e^{-(t-s)/\tau_r} e^{-(t-s')/\tau_r} \langle \pmb\eta(s) \cdot \pmb\eta(s') \rangle, 
\label{eq:v2eq}
\ee
which evaluates to $\langle v^2\rangle_{eq}  = n D/\tau_r$.   This is the reason why in Eq. (\ref{eq:w-0}) we subtracted the quantity 
$n D/\tau_r$ from $\langle v^2\rangle$.

For an unconfined environment ${\bf F}=0$, The expression in Eq. (\ref{eq:v2t}) can be evaluated considering that 
the two random processes are uncorrelated, $\langle {\bf f}(t) \cdot \pmb\eta(t') \rangle = 0$, 
thermal fluctuations are delta correlated, $\langle \pmb \eta(t) \cdot \pmb\eta(t')\rangle = 2n D\delta(t-t')$, 
and the force ${\bf f}$ is exponentially correlated, $\langle {\bf f}(t) \cdot {\bf f}(t')\rangle = \langle f^2\rangle e^{-|t-t'|/\tau}$, 
where $\tau$ is the persistence time.  The evaluated expression is 
\be
\langle v^2\rangle =  \frac{\mu^2 \langle f^2\rangle}{1 + \tau_r/\tau} + \frac{nD}{\tau_r}.
\label{eq:v2tb}
\ee
Inserting this into Eq. (\ref{eq:w-0}) and using expressions for $\langle f^2\rangle$ in Eq. (\ref{eq:f}), the relation yields  
\be
T\Pi \equiv   \langle \dot q\rangle =   \mu \langle f^2\rangle \frac{\tau}{\tau + \tau_r}.
\label{eq:pi-0}
\ee

In the RTP and ABP models the magnitude of ${\bf f}$ is constant, $|{\bf f}| = v_0/\mu$ and we have 
\be
 \langle f^2\rangle =  \frac{v_0^2}{\mu^2}, ~~~~ \text{RTP and ABP}, 
\label{eq:f}
\ee
where $v_0$ represents a swimming velocity that a particle would attain in the overdamped regime in response to 
the force ${\bf f}$.  $v_0$ should not to be confused with the actual velocity $v = |{\bf v}|$.   Inserting this in
Eq. (\ref{eq:pi-0}) results in 
\be
T\Pi \equiv   \langle \dot q\rangle =  \frac{v_0^2}{\mu}  \frac{\tau}{\tau + \tau_r}, ~~~~ \text{RTP and ABP}.
\label{eq:pi}
\ee

The formula for $\langle \dot q\rangle$ is independent of thermal fluctuations.  
For overdamped dynamics, $\tau_r=0$, the heat dissipation becomes independent of $\tau$.  
The inclusion of inertia leads to reduced dissipation as a function of decreasing $\tau$, where in the limit $\tau\to 0$ 
the dissipation vanishes, indicating that a system is in equilibrium.

In the AOUP model, the force ${\bf f}$ evolves as $\tau \dot {\bf f} = -{\bf f} +  \pmb{\eta}_f \mu^{-1}$ 
\cite{Szamel14,Cates21}, where $\pmb\eta_f$ is a delta correlated Gaussian noise, 
$\langle\pmb\eta_f(t) \cdot \pmb\eta_f(t')\rangle = 2n D_f \delta(t-t')$, and $D_{f}$ is the diffusion constant of that process.  
This results in 
\be
 \langle f^2\rangle =  \frac{n D_f}{\tau \mu^2} , ~~~~ \text{AOUP}.  
\label{eq:f-2}
\ee
Inserting this in Eq. (\ref{eq:pi-0}) leads to 
\be
T\Pi \equiv   \langle \dot q\rangle =  \frac{n D_f}{\mu}  \frac{1}{\tau + \tau_r}, ~~~~ \text{AOUP}.
\label{eq:pi-2}
\ee

{The absence of the dependence of $\langle \dot q\rangle$, in the RTP and ABP models, on the system dimension $n$ can be 
traced to the quantity $\langle f^2\rangle$ in Eq. (\ref{eq:f}), and the fact that 
the magnitude of the force ${\bf f}$ is fixed.  In contrast, the same quantity for the AOUP model depends on a system dimension
as a result of Eq. (\ref{eq:f-2}).  In this case, each component of the vector ${\bf f}$ evolves independently.  }

The result in Eq. (\ref{eq:pi}) and in Eq. (\ref{eq:pi-2}) are in agreement with those in \cite{Shankar18} (Table 1 in that reference) 
using different derivation. The results in \cite{Shankar18} are limited to 2D, but the formula in Eq. (\ref{eq:pi}) and Eq. (\ref{eq:pi-2}) 
apply to any system dimension.

\subsection{"Maxwell" distributions of active particles in unconfined environment}

From Eq. (\ref{eq:v2tb}), it can inferred that $\langle v^2\rangle$ is a sum of two contributions:  
\be
\langle v^2\rangle = \langle v^2\rangle_{eq}  +  \langle v^2\rangle_{ac},
\label{eq:v2-sum}
\ee
where $\langle \dots\rangle_{eq}$ is an average over thermal fluctuations (without active motion) and 
$\langle \dots\rangle_{ac}$ is an average due to active motion at zero temperature.  Such a decomposition 
of a second moment is a signature of a distribution generated by convolution,   
$p(v) =  \int d{\bf v}' \, p_{ac}(v') p_{eq}({\bf v}-{\bf v}')$, where 
\be
p_{eq}(v) = \left(\frac{\tau_r}{2\pi D}\right)^{\frac{n}{2}}  e^{-\frac{\tau_r}{2D} v^2},
\label{eq:pmb}
\ee
is the Maxwell-Boltzmann distribution (we recall that $D = \mu k_BT$ and $\tau_r=m\mu$).

Eq. (\ref{eq:pmb}) together with the convolution relation leads to the following distribution in the velocity-space:    
\be
p(v) =  \left(\frac{\tau_r}{2\pi D}\right)^{\frac{n}{2}}  \int_{}^{} d{\bf v}' \, p_{ac}(v') e^{-\frac{\tau_r}{2D} ({\bf v}'-{\bf v})^2}, 
\label{eq:pv-gen}
\ee
where $p_{ac}$ depends on a particular model.

A similar convolution formula has recently been determined for active particles in a harmonic trap \cite{Frydel22c}. 
Since the Langevin equation in (\ref{eq:Langevin-0}) for an unconstrained environment, ${\bf F}=0$, can be 
interpreted as the Langevin equation for an active particle in a harmonic trap in the overdamped regime, we 
expect an analogous convolution relation to apply in this case.   

Convolution arises when the contributing random processes are independent. 
This is the case for the process represented by the sum ${\bf f} + \pmb \eta$.  A more rigorous 
proof of the convolution relation based on Fourier analysis is provided in appendix (\ref{sec:app1}).

To determine the distribution $p(v)$, we still need an expression for $p_{ac}(v)$.  
For the RTP model in dimensions $n=1,2$, $p_{ac}$ is represented by a beta distribution \cite{Frydel22c}
\be
p_{ac}(v) \propto    \left(1 - \frac{v^2}{v_0^2} \right)^{ \frac{n}{2} \frac{\tau_r}{\tau} - 1} ~~~ \text{RTP for $n=1,2$}.  
\label{eq:beta}
\ee
The above distribution vanishes for $v>v_0$ and exhibites a crossover at $\tau/\tau_r = n/2$, 
at which point the distribution changes from convex to concave shape.  In other words, for long persistence 
times, $\tau/\tau_r > n/2$, velocities accumulate near $v=v_0$, and for shorter persistence times, $\tau/\tau_r < n/2$, 
velocities accumulate around $v=0$.

The beta distribution in Eq. (\ref{eq:beta}) is not valid for the RTP model in dimensions $n\ge 3$.  
Likewise, there is no simple closed form formula of $p_{ac}$ for the ABP model in any dimension.  
For the ABP model in 2D, $p(v)$ can be represented as a series solution involving Laguerre polynomials,
see Eq. (11) in \cite{Dhar20}, but there is no such solution at zero temperature or $p_{ac}$.
For those cases where $p_{ac}$ is not available, the 
convolution formula in Eq. (\ref{eq:pv-gen}) can still be used, but the distributions $p_{ac}$ need to be obtained
from a simulation.

For the AOUP model, $p_{ac}$ in any dimension is a Gaussian function \cite{Szamel14} 
\be
p_{ac}(v) 
\propto 
e^{-\frac{v^2}{2} \frac{\tau + \tau_r}{D_f}}  ~~~ \text{AOUP}.  
\label{eq:gaussian1}
\ee
Then application of the convolution relation leads to  
\be
p(v) \propto  e^{ -\frac{v^2}{2} / [\frac{D}{\tau_r} + \frac{D_f}{\tau + \tau_r}] }  ~~~ \text{AOUP}.
\label{eq:gaussian2}
\ee
Note that for $\tau=0$ the distribution in Eq. (\ref{eq:gaussian2}) does not recover the Maxwell distribution
$p_{eq}$ in Eq. (\ref{eq:pmb}), complicating an interpretation of such a system as equilibrium.

\section{Particles in confinement}
\label{sec:sec2}

To analyze active particles in confinement in the underdamped regime, we 
consider the Kramers equations \cite{Kramer} for the evolution of $\rho({\bf r},{\bf v},{\bf f},t)$: 
\ba
\frac{\partial\rho}{\partial t}  &=& -{\bf v} \cdot \bnabla_{r}\rho 
- \frac{1}{\tau_r} \bnabla_v \cdot \left[ \left(  \mu {\bf F}  -  {\bf v}  +  \mu {\bf f} \right) \rho \right]  + \frac{D}{\tau_r^2}  \nabla_v^2 \rho \nonumber\\
&+&\frac{1}{\tau}
\begin{cases}
      \frac{1}{2\pi} \int_0^{2\pi} d\theta_f\, \rho    -  \rho, & \text{RTP} \\
     \frac{\partial^2\rho}{\partial\theta_f^2}, & \text{ABP} \\ 
     \bnabla_{\!\! f} \cdot   {\bf f} \rho   +  \frac{D_f}{\tau \mu^2} \bnabla_f^2 \rho, & \text{AOUP}.
\end{cases}
% \nonumber\\
\label{eq:kramer}
\ea
$\bnabla_{r}$, $\bnabla_v$, and $\bnabla_f$ in Eq. (\ref{eq:kramer}) are the gradient operators with respect to ${\bf r}$, 
${\bf v}$, and ${\bf f}$, and ${\bf F} = - \bnabla V$ is the external time independent force.  The last terms in Eq. (\ref{eq:kramer}) 
govern the evolution of ${\bf f}$ and depend on a specific model.  For the RTP and ABP models, these terms are defined 
for a system in two-dimensions to simplify expressions.

To calculate $\langle v^2\rangle$, or other average quantities of interest, we multiply the stationary Kramers equation,
 $\dot \rho = 0$, by a function $g\equiv g({\bf r},{\bf v},{\bf f})$ and then integrate each term as $\int d{\bf r}\int d{\bf v}\int d{\bf f}$.  
Using integration by parts, the terms of the Kramers equation are represented as average quantities, 
$\langle \dots \rangle = \int d{\bf r}\int d{\bf v}\int d {\bf f}\, \, \, (\dots)$, 
\ba
0 &=& \langle {\bf v}\cdot \bnabla_r g\rangle +
\frac{1}{\tau_r}  \langle  \left(  \mu {\bf F}  -  {\bf v}  +  \mu {\bf f} \right)  \cdot  \bnabla_v g \rangle 
+ \frac{D}{\tau_r^2}  \langle  \nabla_v^2 g \rangle  \nonumber\\
&+&\frac{1}{\tau}
\begin{cases}
     \frac{1}{2\pi}  \langle [\int_0^{2\pi} d\theta_f\, g]  \rangle   -    \langle g \rangle, & \text{RTP} \\
     \langle \frac{\partial^2 g}{\partial\theta_f^2} \rangle, & \text{ABP} \\
     - \langle  {\bf f} \cdot  \bnabla_f g \rangle   +   \frac{D_f}{\tau \mu^2} \langle  \bnabla_f^2 g\rangle, & \text{AOUP}.
\end{cases}
\label{eq:kramer-t}
\ea

Eq. (\ref{eq:kramer-t}) is a template from which various relations can be generated by specifying the function $g$.  
As we are interested in $\langle v^2\rangle$, a logical choice seems to be $g = v^2$, which together with Eq. (\ref{eq:kramer-t}) yields 
\be
%0  =     \frac{\mu}{\tau_r}  \langle {\bf F}\cdot {\bf v}\rangle  
%- \frac{1}{\tau_r}\langle v^2\rangle +  \frac{\mu}{\tau_r}  \langle {\bf f}\cdot{\bf v}\rangle   +  \frac{nD}{\tau_r^2}.  
0  =     \mu  \langle {\bf F}\cdot {\bf v}\rangle  
-  \langle v^2\rangle   +  \mu \langle {\bf f}\cdot{\bf v}\rangle   +  \frac{nD}{\tau_r}.  
\label{eq:g3}
\ee
The above equation looks similar to Eq. (\ref{eq:Langevin-1}) obtained from the Langevin equation.  
Comparing the two equations, we confirm that $ \langle \pmb{\eta}\cdot {\bf v}\rangle =  \frac{nD}{\tau_r}$.

From the above relation it is obvious that for ${\bf f}=0$ we get $0=\langle {\bf F}\cdot {\bf v}\rangle$.  
To confirm that this result holds for nonzero ${\bf f}$ we next use $g=V$.   Eq. (\ref{eq:kramer-t}) in this case 
yields 
\be
0  =  \langle {\bf F}\cdot{\bf v} \rangle, 
\label{eq:g1}
\ee
which shows the above relation to be valid for any stationary system.  We used the relation in Eq. (\ref{eq:g1}) 
to obtain Eq. (\ref{eq:g1a}).  Now we provide a more rigorous proof of it.  

Combining Eq. (\ref{eq:g3}) with Eq. (\ref{eq:g1}) and using the definition of $\langle \dot q\rangle$ in Eq. (\ref{eq:w-0}), 
we get an alternative formulation of the heat dissipation rate, 
\be
\langle \dot q\rangle = \langle {\bf f}\cdot{\bf v}\rangle, %=   \frac{v_0}{\mu} \langle v_{\parallel}\rangle .  
\label{eq:pi2}
\ee
which already was obtained in Eq. (\ref{eq:g1a}).

Other formulations of $\langle \dot q\rangle$ are still possible.  
Inserting $g= {\bf f}\cdot{\bf v}$ into Eq. (\ref{eq:kramer-t}) yields  
\be
0 =  \frac{\mu}{\tau_r} \langle {\bf F} \cdot {\bf f} \rangle  -  \frac{1}{\tau_r} \langle {\bf f}\cdot{\bf v}\rangle 
+ \frac{\mu}{\tau_r} \langle f^2 \rangle  -  \frac{1}{\tau}  \langle {\bf f}\cdot{\bf v}\rangle.  
\label{eq:g4}
\ee
Together with Eq. (\ref{eq:pi2}), Eq. (\ref{eq:g4}) leads to another formulation of the heat dissipation rate:  
\be
\langle \dot q\rangle 
=  \left[ \mu \langle f^2 \rangle
%\frac{v_0^2}{\mu}  
+  \mu \langle {\bf F} \cdot {\bf f} \rangle\right] \frac{\tau}{\tau+\tau_r}.  
\label{eq:pi3}
\ee
The second term can be interpreted as representing contributions of confinement.  The correlations between
vectors ${\bf F}$ and ${\bf f}$ are shown in the next section which treats a harmonic confinement to be negative.  
This means that confinement reduces dissipation.  The negative correlations also explain the accumulation 
of active particles near a trap border.

Another formulation of heat dissipation is obtained by using $g = {\bf r}\cdot{\bf f}$.  Eq. (\ref{eq:kramer-t}) in this case yields  
\be
0  =   \langle {\bf f} \cdot {\bf v} \rangle   -  \frac{1}{\tau}  \langle {\bf f}\cdot{\bf r}\rangle.  
\ee
In combination with Eq. (\ref{eq:pi2}) this leads to 
\be
\langle \dot q\rangle  =  \frac{1}{\tau}  \langle {\bf f}\cdot{\bf r}\rangle, 
\label{eq:pi4}
\ee
the fourth formula of heat dissipation.

All formulations for the heat dissipation rate are listed below:  
\be
T\Pi \equiv   \langle \dot q\rangle =
\begin{cases}
     \frac{1}{\mu} \left [ \langle v^2\rangle - \frac{n D}{\tau_r} \right]  \\
%     \frac{\mu \langle f^2 \rangle}{1+\tau_r/\tau}  +  \frac{\mu}{1+\tau_r/\tau} \langle {\bf F} \cdot {\bf f} \rangle \\
     \left[ \mu \langle f^2 \rangle  +  \mu \langle {\bf F} \cdot {\bf f} \rangle\right] \frac{\tau}{\tau+\tau_r} \\
    \langle {\bf f}\cdot{\bf v}\rangle \\
     \frac{1}{\tau}  \langle {\bf f}\cdot{\bf r}\rangle. 
\end{cases}
\label{eq:pi_all}
\ee
Note that the second and fourth formulas do not directly depend on the velocity and so apply to 
models in the overdamped regime.

The stationary distribution $\rho$ is a function of three vector variables, $\rho\equiv \rho({\bf r},{\bf v},{\bf f})$.  
As shown above, ${\bf f}$ is not an independent variable.  It is on average correlated with other vector variables
$\langle {\bf f}\cdot{\bf r}\rangle$ and $\langle {\bf f}\cdot{\bf v}\rangle$, where the magnitude of correlations is related 
to the dissipation of heat.  To determine correlations between ${\bf r}$ and ${\bf v}$, we insert $g=r^2$ into 
Eq. (\ref{eq:kramer-t}).  The result is 
\be
0  =  \langle {\bf r}\cdot{\bf v} \rangle
\label{eq:g2}
\ee
indicating the absence of correlations for this pair of vectors.

\section{The virial theorem of active particles}

In this section, we are going to use Eq. (\ref{eq:kramer-t}) to obtain a virial theorem for active particles.  
Using $g = {\bf r}\cdot{\bf v}$, Eq. (\ref{eq:kramer-t}) yields 
\be
0  =   \langle v^2 \rangle + \frac{\mu}{\tau_r}  \langle {\bf F}\cdot {\bf r} \rangle   -   \frac{1}{\tau_r} \langle {\bf r}\cdot{\bf v} \rangle   
+  \frac{\mu}{\tau_r}  \langle {\bf f}\cdot{\bf r} \rangle.
\label{eq:g5}
\ee
Together with Eq. (\ref{eq:pi4}), Eq. (\ref{eq:w-0}), and Eq. (\ref{eq:g2}, the above equation can be expressed as
\be
-\langle {\bf F}\cdot {\bf r} \rangle  
=   2\langle E_k \rangle + \tau  \langle \dot q \rangle, 
\label{eq:virial}
\ee  
where $\langle E_k \rangle = \langle m v^2/2\rangle$ is the average kinetic energy where we used $\tau_r=m\mu$ 
to eliminate $\tau_r$.  The relation in Eq. (\ref{eq:virial}) is the virial theorem for active particles.  
Since at equilibrium the virial theorem is $ -\langle {\bf F}\cdot {\bf r} \rangle_{} = 2 \langle E_k \rangle_{}$, 
the term $\tau  \langle \dot q \rangle$ in Eq. (\ref{eq:virial}) 
can be regarded as a measure of violation of the virial relation.  

The virial theorem in Eq. (\ref{eq:virial}) can be expressed differently by using the fourth formulation in Eq. (\ref{eq:pi_all}), 
leading to 
\be
-\langle ({\bf F}+ {\bf f}) \cdot {\bf r} \rangle  =   2\langle E_k \rangle.  
\ee
Given that ${\bf f}$ is an external force, the formulation above appears to conform with the original virial theorem.

A virial relation is often considered for the following external potential $V = Kr^s / s$.  In this case 
${\bf F} \cdot {\bf r} = -s V$ and Eq. (\ref{eq:virial}) becomes 
\be
s \langle V \rangle =  2\langle E_k \rangle  +  \tau \langle \dot q\rangle.  
\label{eq:virial2}
\ee
Active particles for this class of potentials have been studied in \cite{Dhar19}.  

The virial theorem has been previously derived specifically for the AOUP model and using other 
derivation techniques, see Eq. (28) in \cite{Lorenzo21}.  The formulation in this work
extends this result to all active particle models.

\section{Particles in a harmonic trap} 
\label{sec:sec3}

In this section we focus on a specific confining potential, a harmonic trap, resulting in the linear external force 
${\bf F} = -K{\bf r}$.  To obtain an expression for the dissipation of heat, we use Eq. (\ref{eq:pi3}).   And to 
obtain an expression for $\langle {\bf F}\cdot{\bf f} \rangle$, we substitute into Eq. (\ref{eq:kramer-t}) 
$g = {\bf F}\cdot {\bf f}$.  This yields
\be
0 =  \langle {\bf v} \cdot  \bnabla_r ({\bf F}\cdot{\bf f}) \rangle - \frac{1}{\tau} \langle {\bf F}\cdot{\bf f} \rangle. 
\label{eq:R5}
\ee
For ${\bf F} = -K{\bf r}$ the first term becomes 
$\langle {\bf v} \cdot  \bnabla_r ({\bf F}\cdot{\bf f}) \rangle =  - K\langle {\bf v} \cdot {\bf f} \rangle$, 
then using Eq. (\ref{eq:pi2}), Eq. (\ref{eq:R5}) can be written as 
\be
 \langle {\bf F}\cdot{\bf f} \rangle = - \tau K \langle \dot q\rangle.  
\label{eq:R6}
\ee
Note that the correlations between the vectors ${\bf F}$ and ${\bf f}$ are 
demonstrated to be negative.  

Substituting this result into Eq. (\ref{eq:pi3}), 
the heat dissipation formula for a harmonic trap becomes 
\be
T\Pi \equiv   \langle \dot q\rangle =
     \mu \langle f^2 \rangle  \frac{\tau}{\tau + \tau^2 \mu K  + \tau_r}.  
\label{eq:pi_Ka}
\ee
Then using Eq. (\ref{eq:f}), we get the dissipation of heat for the RTP and ABP models:  
\be
T\Pi \equiv   \langle \dot q\rangle =
     \frac{v_0^2}{\mu}  \frac{\tau}{\tau + \tau^2 \mu K  + \tau_r}, ~~~~ \text{RTP and ABP}.
\label{eq:pi_K}
\ee
The entropy production rate for RTP particles in a harmonic trap have previously been obtained for 
the overdamped regime, or $\tau_r=0$, in \cite{Pruessner21,Frydel22a}.  The formulas in Eq. (\ref{eq:pi_K})
recover these results by setting $\tau_r$ to zero.  

Similar to the case of an unconfined system, $\langle \dot q\rangle$ is independent of temperature.  The role of confinement, 
regulated by the parameter $K$, is to reduce a dissipation effect.  Because the term that depends on confinement is 
$\tau^2 \mu K$, the effects of confinement becomes negligible as $\tau$ becomes small and the dissipation is
similar to that in an unconfined environment.

\begin{figure}[hhhh] 
 \begin{center}
 \begin{tabular}{rrrr}
\includegraphics[height=0.21\textwidth,width=0.29\textwidth]{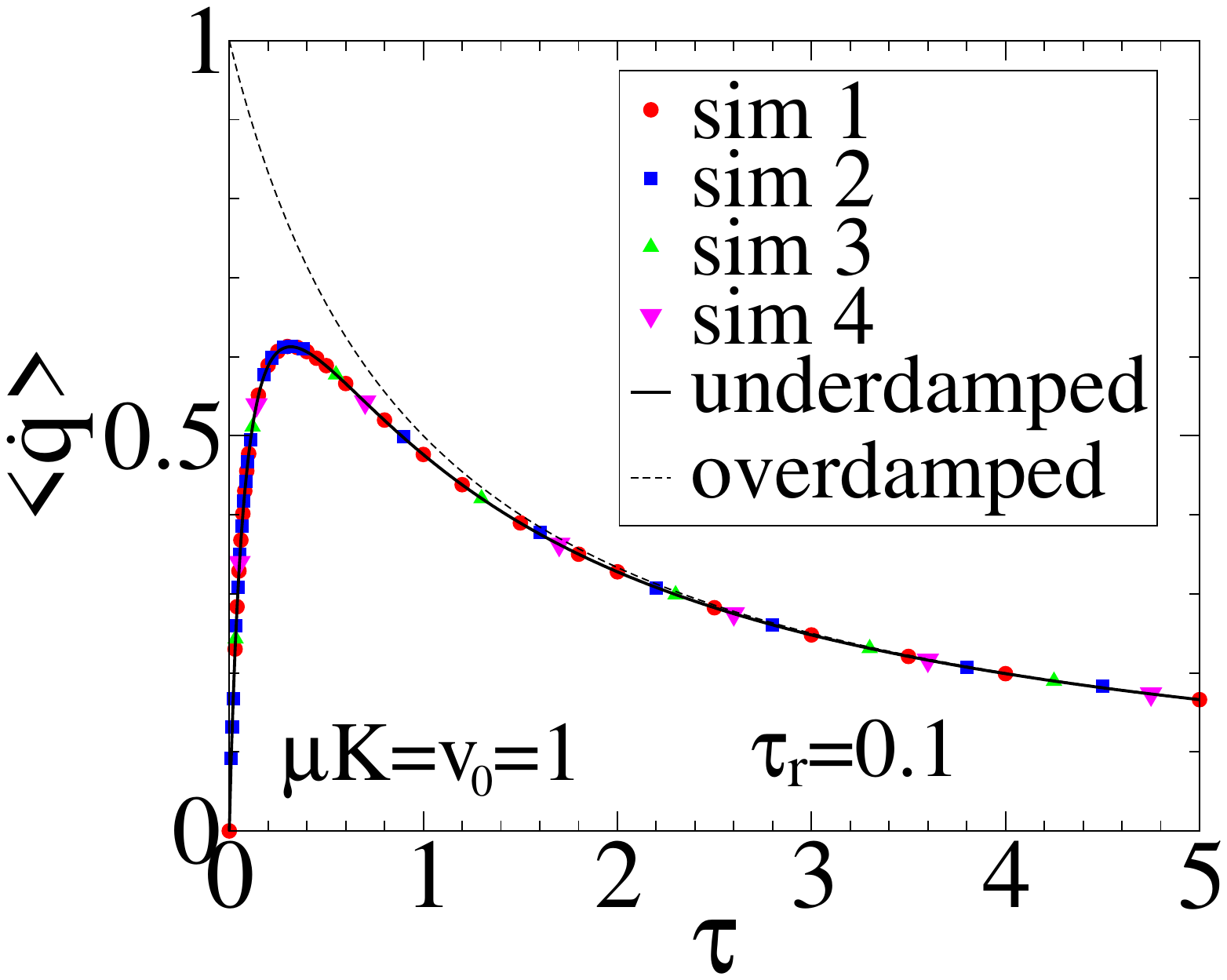}  
 \end{tabular}
 \end{center} 
\caption{The heat dissipation rate of self-propelled particles in a harmonic trap as a function of $\tau$.
Simulation data points come from four systems.  Sim-1 is for RTP particles in $2D$ and $D=0$, sim-2
is for RTP particles in 1D and $D=0.1$, sim-3 for ABP particles in 2D and $D=0.2$, and sim-4 for ABP 
particles in 3D and $D=0.2$.   All points collapse onto the curve given by the formula in Eq. (\ref{eq:pi_K}).  
The dashed line is $\langle \dot q\rangle$ for the overdamped regime.  Maximum dissipation occurs at
$\tau_{max} = \sqrt{m/K}$, for the specific parameters of the plot corresponding to $\tau_{max} = \sqrt{0.1}$.}
\label{fig:fig3a} 
\end{figure}

In Fig. (\ref{fig:fig3a}) we plot simulation data points for $\langle \dot q\rangle$ as a function of $\tau$
for the RTP and ABP models, different system dimensions, and different temperatures.   According to 
Eq. (\ref{eq:pi_K}), all points are expected to collapse onto a single curve, which is confirmed by the 
results.  To calculate $\langle \dot q \rangle$, we can use any of the formulas in Eq. (\ref{eq:pi_all}).  
All the formulas were tested and shown to yield the same results.  
 %%%%%%%%%%%%%%%%%%%%%%
\graphicspath{{figures/}}
%%%%%%%%%%%%%%%%%%%%%%%

The simulation data points are compared against $\langle \dot q\rangle$ for overdamped dynamics, 
obtained from Eq. (\ref{eq:pi_K}) with $\tau_r=0$.  
The two curves deviate in the region $\tau\lesssim \tau_r$.
As $\langle \dot q\rangle$ quantifies a distance from equilibrium, this leads to different interpretation
of the same system when analyzed using different theoretical frameworks.   

{As seen in Fig. (\ref{fig:fig3a}), the maximal dissipation of heat for the underdamped regime occurs at 
non-zero persistence time, corresponding 
to $\frac{d\langle \dot q\rangle}{d\tau} = 0$ and that can be calculated using Eq. (\ref{eq:pi_K}), yielding 
\be
\tau_{max} = \sqrt{\frac{m}{K}} = \omega^{-1},
\ee
where $\omega$ is the natural frequency of an oscillator.  
This result might be expected based on what is known about driven harmonic oscillators.  }

Using Eq. (\ref{eq:pi_Ka}) and Eq. (\ref{eq:f-2}), we get 
the heat dissipation formula for AOUP particles in harmonic trap:  
\be
T\Pi \equiv   \langle \dot q\rangle =
%\begin{cases}
%     \frac{v_0^2}{\mu}  \frac{\tau}{\tau + \tau^2 \mu K  + \tau_r}, ~~~~ \text{RTP and ABP}.
     \frac{n D_f}{\mu}   \frac{1}{\tau + \tau^2 \mu K  + \tau_r}, ~~~~~ \text{AOUP}.
%\end{cases}
\label{eq:pi_K-2}
\ee

%\textcolor{red}
{As in the case of unconfined system, the expressions for $\langle \dot q\rangle$ for the RTP and ABP models
in Eq. (\ref{eq:pi_K}) does not depend on a system dimension $n$.  This has to do with the fact that the magnitude of 
the force vector ${\bf f}$ is fixed.  In contrast, Eq. (\ref{eq:pi_K-2}) shows
dependence on a system dimension since the vector components of ${\bf f}$ evolve independently, see
Eq. (\ref{eq:f-2}).  }

\subsection{"Maxwell" distributions of active particles in a harmonic trap}

Because Eq. (\ref{eq:pi_Ka}) is independent of temperature, this implies that the variance can be written as 
$\langle v^2\rangle = \langle v^2\rangle_{eq} +  \langle v^2\rangle_{ac}$ which, in trun, implies 
that a distribution in $v$-space obeys the convolution formula $p(v) =  \int d{\bf v}' \, p_{ac}(v') p_{eq}({\bf v}-{\bf v}')$ 
as for the case of unconfined environment in Eq. (\ref{eq:pv-gen}). 

The validity of the convolution formula implies that the independence of the random processes ${\bf f}$ and 
$\pmb{\eta}$ is not violated by the presence of a linear force of a harmonic trap.  In any other type of external 
potential, the independence of the two processes can no longer be assumed.  
A more technical demonstration of the validity of the convolution relation for particles in a harmonic trap is 
given in appendix (\ref{sec:app2}) using Fourier transform analysis.

For the RTP and ABP models, there are no exact expressions for $p_{ac}$ for particles in a harmonic trap, and 
these distributions need to be obtained from a simulation.   For the AOUP model, $p_{ac}(v)$ is a Gaussian function 
\cite{Szamel14}.  Given the value of $\langle v^2\rangle_{ac}$ in Eq. (\ref{eq:pi_K-2}), we know that this 
function must be given by 
\be
p_{ac} (v) \propto e^{-\frac{\tau + \tau_r + \tau^2\mu K}{D_f} \frac{v^2}{2}}, ~~~~~~ \text{AOUP}.  
\ee
Using the convolution formula in Eq. (\ref{eq:pv-gen}) we then get 
\be
p_{}(v) \propto  e^{ -\frac{v^2}{2} / \big[\frac{D}{\tau_r} + \frac{D_f}{\tau + \tau_r + \tau^2\mu K}\big] },  ~~~ \text{AOUP}.  
\label{eq:gaussian4}
\ee

\section{RTP model in 1D box}
\label{sec:sec5}

A frequently studies system of active particles are RTP particles in 1D box.  
In the overdamped regime, the model has been frequently studies and is well understood 
\cite{Schnitzer93, Angelani17, Dhar18, Dhar19,Basu20,Razin20, Frydel22b}.  For underdamped dynamics, 
the system is governed by the Kramers equation resulting in two coupled differential equations:  
\ba
&&\dot p_+ = -v\frac{\partial p_+}{\partial x} 
+ \frac{1}{\tau_r} \frac{\partial }{\partial v}\left[ \left(v  - v_0\right) p_+ \right]  + \frac{D}{\tau_r^2}  \frac{\partial^2 p_+}{\partial v^2} -   \frac{1}{2} \frac{1}{\tau} (p_+  -  p_-)
\nonumber\\
&&\dot p_-  = -v\frac{\partial p_-}{\partial x} 
+ \frac{1}{\tau_r} \frac{\partial }{\partial v}\left[ \left(v  + v_0\right) p_- \right]  + \frac{D}{\tau_r^2}  \frac{\partial^2 p_-}{\partial v^2} +  \frac{1}{2} \frac{1}{\tau} (p_+  -  p_-),
\nonumber\\
\label{eq:kramer-rtp1d}
\ea
where $p_+(x,v,t)$ and $p_-(x,v,t)$ are the distributions for particles subject to a forward and backward force.  
The entropy production for this model in the overdamped regime, or $\tau_r=0$, has been obtained in \cite{Razin20}.  
For the walls located at $x=\pm h$ it reads
\be
%\langle \dot q\rangle 
T\Pi = \langle \dot q\rangle 
= \frac{v_0^2}{\mu}  
%\frac{1}{1+\tau_r/\tau}
\left[ \frac{D}{v_0^2\tau }  \frac{ \cosh(kh) - \frac{\sinh(kh)}{kh}  }{ \frac{D}{v_0^2\tau}\cosh(kh) + \frac{\sinh(kh)}{kh} } \right], 
\label{eq:w-rtp1d}
\ee
where
$
k = \frac{v_0}{D}\sqrt{1+\frac{D}{v_0^2\tau}}. 
$
We can re-derive this result using one of the formulas in Eq. (\ref{eq:pi_all}), which further demonstrates
the accuracy of those formulas.

To calculate $\langle \dot q\rangle$ for the underdamped regime, we use the last 
formula in Eq. (\ref{eq:pi_all}), which for the present model becomes 
\be
T\Pi  
= \frac{v_0}{\tau\mu}  \left[ \langle x\rangle_{+}   -   \langle x\rangle_{-} \right], 
\label{eq:a}
\ee
where $\langle \dots \rangle_{\pm} = \int_{-h}^h dx\, (\dots) p_{\pm}$. 

\begin{figure}[hhhh] 
 \begin{center}
 \begin{tabular}{rrrr}
\includegraphics[height=0.21\textwidth,width=0.27\textwidth]{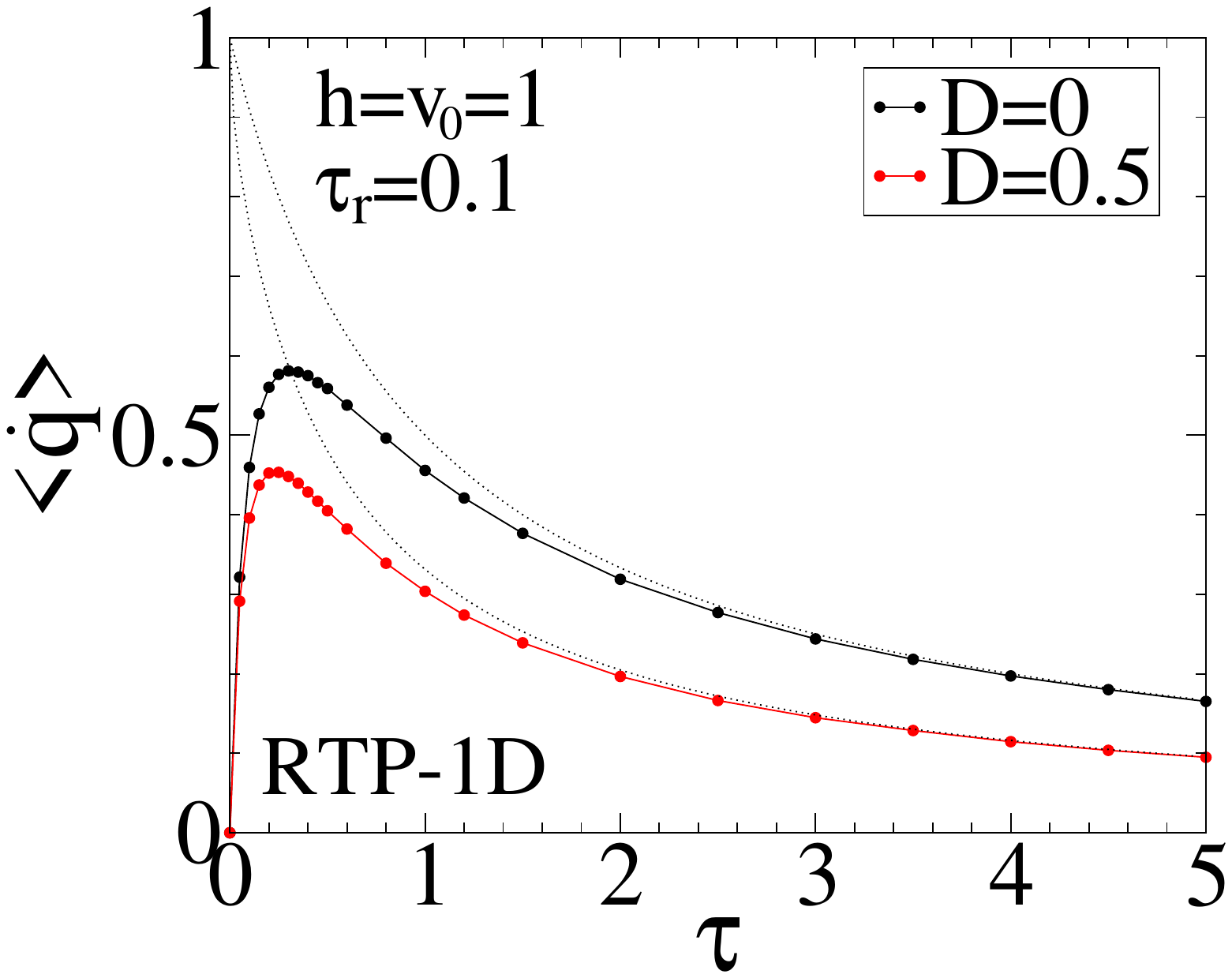}  
 \end{tabular}
 \end{center} 
\caption{The heat dissipation rate as a function of $\tau$ for the RTP particles in 1D confined between two walls
located at $x=\pm h$.   
Dashed lines represent the exact expression in Eq. (\ref{eq:w-rtp1d}) for the overdamped regime, 
and the data points for a finite $\tau_r$ are obtained from the simulation.}
\label{fig:fig3b} 
\end{figure}

Since distributions $p_{\pm}$ in Eq. (\ref{eq:kramer-rtp1d}) cannot be solved exactly, we use simulation to 
evaluate the expression in Eq. (\ref{eq:a}).  
%\textcolor{red}
{Interaction with the walls is accounted for as follows:    
each time a particles reaches a wall, its instantaneous velocity changes as $v(t)\to -v(t)$.  }
The data points for $\langle \dot q\rangle$ obtained from a simulation   
are plotted in Fig. (\ref{fig:fig3b}).  For $\tau>2$, the simulation results agree with the exact formula in 
Eq. (\ref{eq:w-rtp1d}) for the overdamped regime.  
%Deviation between  overdamped and underdamped dynamics become clear for small values of $\tau$.    
%%%%%%%%%%%%%%%%%%%%%%
\graphicspath{{figures/}}
%%%%%%%%%%%%%%%%%%%%%%%

For particles in unconfined environment and ina harmonic trap the dissipation of heat was found to be  
independent of temperature.  In contrast, for particles in 1D box we observe strong temperature dependence, 
and the indication that thermal fluctuations reduce dissipation of heat (the difference between the black 
and red curves).  

%\textcolor{red}{
Interactions of a particle with a wall are analogous to a bouncing ball in the air (within the time duration 
in which the force $f$ does not change its direction).  Due to air friction, 
a maximum height at each bounce is lower and the overall velocity reduced.  
Since interactions with the walls reduce velocity of a particle 
(and, therefore, the dissipation of heat), 
increased thermal fluctuations can be seen as contributing to greater number of collisions with the walls.
In fact, any type of confinement, other than a harmonic trap, will show a similar reduction of 
heat dissipation with increased temperature.  This was verified by simulations carried out 
for different trapping potentials of the form $U \propto r^s$.

\begin{figure}[hhhh] 
 \begin{center}
 \begin{tabular}{rrrr}
\includegraphics[height=0.21\textwidth,width=0.27\textwidth]{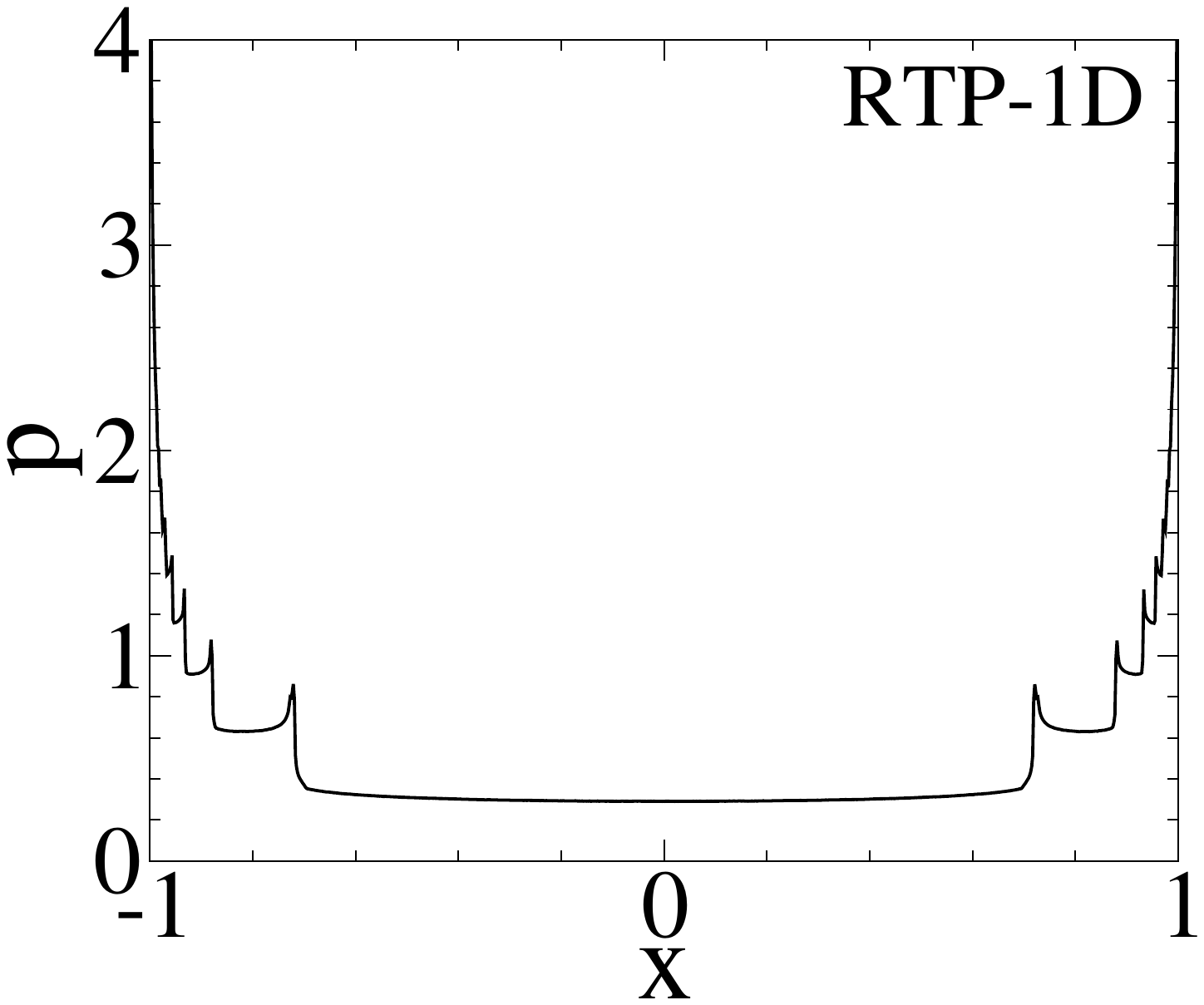}  
 \end{tabular}
 \end{center} 
\caption{A stationary distribution of RTP particles inside a 1D box at zero temperature.  The other 
system parameters are $h=\tau=\tau_r=1$. The distribution is obtained from a simulation.  }
\label{fig:fig3c} 
\end{figure}

There is another interesting feature that arises as a consequence of underdamped dynamics.  Within the overdamped regime 
and at zero temperature, a fraction of particles becomes adsorbed onto confining walls due to the combination of overdamped 
dynamics and finite persistence time \cite{rudi18}.  Particles that are not adsorbed are uniformly distributed within the box \cite{Frydel22b}. 
Within the underdamped dynamics, particles coming against a wall no longer become immediately adsorbed.  Instead, they 
are reflected from it, and if the persistence time is sufficiently long, they change direction and come toward the wall again, 
creating a behavior analogous to bouncing of a ball.  As a result, stationary distributions inside the box 
develop interesting structure shown in Fig. (\ref{fig:fig3c}).  This structure has nothing to do with the size of particles 
and is entirely a dynamical phenomenon.  
%%%%%%%%%%%%%%%%%%%%%%
\graphicspath{{figures/}}
%%%%%%%%%%%%%%%%%%%%%%%

\section{Conclusion}
\label{sec:5}

In this work we calculate the entropy production rate in the underdamped regime for all canonical models of 
active particles in all dimensions and for different confinements.  To calculate $\Pi$, we explore the link between the 
entropy production rate and the dissipation of heat, which results in Eq. (\ref{eq:w-0}) and 
from which alternative formulations are obtained in Eq. (\ref{eq:pi_all}) by analyzing the Kramers equation.

Exact results are obtained for particles in a harmonic trap and in unconfined environment.  
In those two cases, the two random processes $\pmb{\eta}$ and ${\bf f}$ remain independent which 
permits us to represent the "Maxwell" distribution $p(v)$ to be represented as a convolution relation.  
For a harmonic potential, the maximum dissipation of heat occurs for $\tau=\sqrt{m/K}$.  The  
dissipation of heat vanishes for $\tau=0$ and in the limit $\tau\to\infty$.  

For other forms of confinement the two random processes $\pmb{\eta}$ and ${\bf f}$ are generally not independent
and as a result the dissipation of heat is found to be reduced as a result of thermal fluctuations.

%------------------------------------------------
\begin{acknowledgments}
D.F. acknowledges financial support from FONDECYT through grant number 1201192.  
\end{acknowledgments}

\section{DATA AVAILABILITY}
The data that support the findings of this study are available from the corresponding author upon 
reasonable request.

\appendix
\section{Convolution formulation for an unconfined environment}
\label{sec:app1}

The convolution formulation of the distribution $p(v)$ can be demonstrated by analyzing the stationary 
Fokker-Planck equation for the evolution of 
$\rho({\bf v},{\bf f})$:
\ba
0  &=&  \frac{1}{\tau_r} \bnabla_v \cdot \left[ \left(  {\bf v}  -  \mu {\bf f}  \right) \rho \right]  + \frac{D}{\tau_r^2}  \nabla_v^2 \rho \nonumber\\
&+&\frac{1}{\tau}
\begin{cases}
      \frac{1}{2\pi} \int_0^{2\pi} d\theta_f\, \rho    -  \rho, & \text{RTP} \\
     \frac{\partial^2\rho}{\partial\theta_f^2}, & \text{ABP} \\
     \bnabla_{\!\! f} \cdot   {\bf f} \rho   +  \frac{D_f}{\tau \mu^2} \bnabla_f^2 \rho, & \text{AOUP},
\end{cases}
\label{eq:app1}
\ea
where the last term determines the evolution of ${\bf f}$ and depends on the model.  For convenience, these terms
are written specifically for 2D where ${\bf f}=\mu^{-1} v_0{\bf n}$ (the unit orientation vector 
${\bf n} = (\cos\theta_f,\sin\theta_f)$ depends only on the angle $\theta_f$).  For the AOUP model, $\bnabla_f$ is the gradient 
operator with respect to ${\bf f}$.  

Taking the Fourier transform of Eq. (\ref{eq:app1}) with respect to velocity yields 
\ba
0  &=& 
 -  \frac{1}{\tau_r} [{\bf k}  \cdot  \bnabla_k c]  -   \frac{\mu}{\tau_r} [i {\bf k}  \cdot  {\bf f}]  c  -  \frac{D k^2}{\tau_r^2}   c \nonumber\\ 
&+&
\frac{1}{\tau}
\begin{cases}
      \frac{1}{2\pi} \int_0^{2\pi} d\theta_f\, c    -  c, & \text{RTP} \\
     \frac{\partial^2c}{\partial\theta_f^2}, & \text{ABP} \\
     \bnabla_{\!\! f} \cdot   {\bf f} c   +  \frac{D_f}{\tau \mu^2} \bnabla_f^2 c, & \text{AOUP},
\end{cases}
\label{eq:fp-trans}
\ea
where $c({\bf k},{\bf f}) = \int d{\bf v}\, \rho({\bf v},{\bf f}) \, e^{-i{\bf k}\cdot {\bf v}}$ and 
$c_{eq} \propto e^{-D k^2 / 2\tau_r}$ is the transformed Maxwell distribution.  
The same equation at zero temperature reads
\ba
0  &=& 
 -  \frac{1}{\tau_r} [{\bf k}  \cdot  \bnabla_k c_{ac}]  -   \frac{\mu}{\tau_r} [i {\bf k}  \cdot  {\bf f}] c_{ac}\nonumber\\ 
&+&
\frac{1}{\tau}
\begin{cases}
      \frac{1}{2\pi} \int_0^{2\pi} d\theta_f\, c_{ac}    -  c_{ac}, & \text{RTP} \\
     \frac{\partial^2c_{ac}}{\partial\theta_f^2}, & \text{ABP} \\
     \bnabla_{\!\! f} \cdot   {\bf f} c   +  \frac{D_f}{\tau \mu^2} \bnabla_f^2 c, & \text{AOUP}. 
\end{cases}
\label{eq:fp-trans0}
\ea

In the Fourier space, the convolution formula in Eq. (\ref{eq:pv-gen}) can be written as 
\be
c({\bf k},{\bf f}) \propto c_{ac}({\bf k},{\bf f}) e^{-D k^2 / 2\tau_r}.  
\label{eq:ansatz}
\ee
Substituting Eq. (\ref{eq:ansatz}) into Eq. (\ref{eq:fp-trans}) 
splits the first term of that equation into two parts, 
\be
\frac{1}{\tau_r}{\bf k} \cdot \bnabla_k c =   \left( \frac{1}{\tau_r} {\bf k} \cdot \bnabla_k c_{ac}    -   \frac{D k^2}{\tau_r^2}   c_{ac} \right) e^{-{Dk^2}/{2\tau_r}}. 
\label{eq:dum}
\ee
where the first term cancels the third term in that equation, recovering Eq. (\ref{eq:fp-trans0}). 

The solution in Eq. (\ref{eq:ansatz}) not only proves the convolution formula in Eq. (\ref{eq:pv-gen}), but it implies a stronger claim, 
$$
\rho({\bf v},{\bf f}) =  \left(\frac{\tau_r}{2\pi D}\right)^{\frac{n}{2}}  \int_{}^{} d{\bf v}' \, \rho_{ac}({\bf v},{\bf f}) e^{-\frac{\tau_r}{2D} ({\bf v}'-{\bf v})^2}, 
$$
where $\rho_{ac}$ is the distribution for $D=0$.  Consequently, the effect of temperature on all the distributions $\rho_{ac}$
is the same.  Since $p(v) = \int d{\bf f}\, \rho({\bf v},{\bf f})$, the above relation proves the convolution in Eq. (\ref{eq:pv-gen}).  
The proof is valid to all active particle models in unconfined environment.

\section{Convolution formulation for a harmonic trap}
\label{sec:app2}

In this section we demonstrate that the distribution $p(v)$ for active particles in a harmonic trap 
can be represented as a convolution relation between $p_{ac}(v)$ and $p_{eq}(v)$.  
The stationary Kramers equation for particles in a harmonic trap is 
\be
0  = -{\bf v} \cdot \bnabla_{r}\rho 
+ \frac{1}{\tau_r} \bnabla_v \cdot \left[ \left(  \mu K {\bf r}  +  {\bf v}  -  \mu {\bf f} \right) \rho \right]  + \frac{D}{\tau_r^2}  \nabla_v^2 \rho + \dots,
\label{eq:k1}
\ee
where the terms for the evolution of ${\bf f}$, that do not play role in the proof, are ignored.  We next take the
Fourier transform with respect to ${\bf r}$ and ${\bf v}$ that yields 
\be
0  =  {\bf q} \cdot \bnabla_k c  - \frac{ \mu K }{\tau_r}  {\bf k}  \cdot  \bnabla_q  c  
-\frac{1}{\tau_r} {\bf k}  \cdot  \bnabla_k c  -   \frac{\mu}{\tau_r} [i {\bf k}  \cdot  {\bf f}]  c  -  \frac{D k^2}{\tau_r^2} c  + \dots
\label{eq:k2}
\ee
where 
$
c({\bf q},{\bf k},{\bf f}) = \int d{\bf r} \int d{\bf v}\, \rho({\bf r},{\bf v},{\bf f}) \, e^{-i{\bf q}\cdot {\bf r}}e^{-i{\bf k}\cdot {\bf v}}.  
$ 
We next assume the following solution 
\be
c  =  c_{ac}  e^{-D k^2 / 2\tau_r}  e^{-D q^2 / 2\mu K},
\label{eq:k3}
\ee
where $\propto e^{-D k^2 / 2\tau_r}$ is a Fourier transformed Maxwell distribution in $v$-space 
and $\propto e^{-D q^2 / 2\mu K}$ is a Fourier transformed equilibrium distribution in positional space.  
It turns out that if we substitute this function into Eq. (\ref{eq:k2}), we will eliminate all 
the terms that depend on $D$ and recover
\be
0  =  {\bf q} \cdot \bnabla_k c_{ac}  - \frac{ \mu K }{\tau_r}  {\bf k}  \cdot  \bnabla_q  c_{ac} 
-\frac{1}{\tau_r} {\bf k}  \cdot  \bnabla_k c_{ac}  -   \frac{\mu}{\tau_r} [i {\bf k}  \cdot  {\bf f}]  c_{ac}  + \dots, 
\label{eq:k4}
\ee
which is the same as Eq. (\ref{eq:k2}) but for $D=0$.  This means that Eq. (\ref{eq:k3}) is a correct solution.
In physical space, this implies the following convolution:  
\be
\rho({\bf r},{\bf v},{\bf f}) \propto  \int d{\bf r}' \int d{\bf v}'  \, \,    \rho_{ac}({\bf r}',{\bf v}',{\bf f}) e^{-\frac{\mu K}{2D} ({\bf r}-{\bf r}')^2}  e^{-\frac{\tau_r}{2D} ({\bf v}-{\bf v}')^2}.  
\ee
After integrating over ${\bf r}$ and ${\bf f}$ we recover the convolution in Eq. (\ref{eq:pv-gen}).

%------------------------------------------------
% References
%------------------------------------------------

%------------------------------------------------


\begin{thebibliography}{99}



%%%  Entropy production  %%%
\bibitem{Rosen60}
G. Rosen, 
{\sl Entropy Production and Pressure Waves}, 
Phys. Fluids {\bf 3}, 188 (1960);

\bibitem{Schnakenberg76}
J. Schnakenberg, 
{\sl Network theory of microscopic and macroscopic behavior of master equation systems},
Rev. Mod. Phys. {\bf 48}, 571 (1976).

\bibitem{Tome06}
T. Tom\'e, 
{\sl Entropy production in non-equilibrium systems described by a Fokker-Planck equation}, 
Braz. J. Phys. {\bf 36}, 1285 (2006).

\bibitem{Tome12}
T. Tom\'e and M. J. de Oliveira,
{\sl Entropy Production in Nonequilibrium Systems at Stationary States},
Phys. Rev. Lett. {\bf 108}, 020601 (2012)

%\bibitem{Seifert05}
%U. Seifert, 
%{\sl Entropy Production along a Stochastic Trajectory and an Integral Fluctuation Theorem}, 
%Phys. Rev. Lett {\bf 95}, 040602 (2005).

\bibitem{Cates16}
\'E. Fodor, C. Nardini, M. E. Cates, J. Tailleur, P. Visco, and F. van Wijland 
{\sl How Far from Equilibrium Is Active Matter?},  
Phys. Rev. Lett. {\bf 117}, 038103 (2016).

\bibitem{Shankar18}
S. Shankar and M. C. Marchetti, 
{\sl Hidden entropy production and work fluctuations in an ideal active gas}, 
Phys. Rev. E 98, 020604(R) (2018).

\bibitem{Schmidt19}
E. G. Idrisov and T. L. Schmidt, 
{\sl Entropy production in one-dimensional quantum fluids}, 
Phys. Rev. B {\bf 100}, 165404 (2019).

\bibitem{Pruessner20}
L. Cocconi, R. Garcia-Millan, Z. Zhen, B. Buturca, G. Pruessner,
{\sl Entropy production in exactly solvable systems}, 
Entropy {\sl 22}, 1252 (2020).

\bibitem{Razin20}
N. Razin, 
{\sl Entropy production of an active particle in a box}, 
Phys. Rev. E (R) {\bf 102}, 030103(R) (2020).  

\bibitem{Pruessner21}
R. Garcia-Millan and G. Pruessner 
{\sl Run-and-tumble motion in a harmonic potential: field theory and entropy production}, 
J. Stat. Mech. 063203 (2021).  

\bibitem{Leonid10}
L. M. Martyushev
{\sl The maximum entropy production principle: two basic questions}, 
Phil. Trans. R. Soc. B {\bf 365}, 1333 (2010) .




%%%  Underdamped regime  %%%
\bibitem{Frydel22a}
D. Frydel,
{\sl Intuitive view of entropy production of ideal run-and-tumble particles}, 
Phys. Rev. E {\bf 105}, 034113 (2022).





%%%  Underdamped regime  %%%
\bibitem{Lowen19}
L\"owen, H. 
{\sl Inertial effects of self-propelled particles: From active Brownian to active Langevin motion}, 
J. Chem. Phys. {\bf 152}, 040901 (2020). 

\bibitem{Sandoval20}
L. L. Gutierrez-Martinez, M. Sandoval, 
{\sl  Inertial effects on trapped active matter},
J. Chem. Phys. {\bf 153}, 044906. (2020). 

\bibitem{Lowen22}
G. H. P. Nguyen, R. Wittmann, and H. L\"owen, 
{\sl Active Ornstein–Uhlenbeck model for self-propelled particles with inertia},
J. Phys.: Condens. Matter {\bf 34}, 035101 (2022).

\bibitem{Morfill09}
G. E. Morfill and A. V. Ivlev, 
{\sl Complex plasmas: An interdisciplinary research field},  
Rev. Mod. Phys. {\bf 81}, 1353 (2009).

\bibitem{Weber13}
C. A. Weber, T. Hanke, J. Deseigne, S. L\'eonard, O. Dauchot, E. Frey, and
H. Chat\'e, 
{\sl Long-range ordering of vibrated polar disks}, 
Phys. Rev. Lett. {\bf 110}, 208001 (2013).

\bibitem{Kim16}
H. Mukundarajan, T. C. Bardon, D. H. Kim, and M. Prakash, 
{\sl Surface tension dominates insect flight on fluid interfaces}, 
J. Exp. Biol. {\bf 219}, 752 (2016).






%%%  Entropy production general papers %%%
\bibitem{Sekimoto98}
K. Sekimoto, 
{\sl Langevin Equation and Thermodynamics}, 
Progress of Theoretical Physics Supplement, {\bf 130}, 17 (1998).  

\bibitem{Landi21}
G. T. Landi and M. Paternostro, 
{\sl Irreversible entropy production: From classical to quantum}, 
Rev. Mod. Phys. {\bf 93}, 035008 (2021). 

\bibitem{Sasa05}
T. Harada and S. Sasa, 
{\sl Equality Connecting Energy Dissipation with a Violation of the Fluctuation-Response Relation},
Phys. Rev. Lett. {\bf 95}, 130602 (2005).







%%%  PI as heat dissipation  %%%
\bibitem{Maes09}
Maes C, Netocn\'y K and Shergelashvili B 2009
{\sl A Selection of Nonequilibrium Issues in Methods of Contemporary Mathematical Statistical Physics}, 
ed K Roman (Berlin: Springer) pp 247–306.  

\bibitem{Maes21}
T. Banerjee and C. Maes, 
{\sl Active gating: rocking diffusion channels},
J. Phys. A: Math. Theor. {\bf 54}, 025004 (2021).



\bibitem{Frydel21a}
D. Frydel, 
{\sl Stationary distributions of propelled particles as a system with quenched disorder}, 
Phys. Rev. E {\bf 103},  052603 (2021).

\bibitem{Frydel22c}
D. Frydel,
{\sl Positing the problem of stationary distributions of active particles as third-order differential equation}, 
Phys. Rev. E {\bf 106}, 024121,  (2022).




\bibitem{Celani12}
A. Celani, S. Bo, R. Eichhorn, and E. Aurell,
{\sl Anomalous Thermodynamics at the Microscale},
Phys. Rev. Lett. {\bf 109}, 260603 (2012).  


\bibitem{Nakayama13}
K. Kawaguchi and Y. Nakayama, 
{\sl Fluctuation theorem for hidden entropy production}, 
Phys. Rev. E {\bf 88}, 022147 (2013).










\bibitem{Chun15}
H.-M. Chun and J. D. Noh, 
{\sl Hidden entropy production by fast variables}, 
Phys. Rev. E {\bf 91}, 052128 (2015).



\bibitem{Lorenzo21}
Lorenzo Caprini,
{\sl Generalized fluctuation–dissipation relations holding in non-equilibrium dynamics},
J. Stat. Mech. 063202 (2021).











\bibitem{Prigogine55}
I. Prigogine, 
{\sl Introduction to Thermodynamics of Irreversible Processes}, 
(Thomas, Springfield, 1955).

\bibitem{Groot62}
S. R. de Groot and P. Mazur, 
{\sl Non-Equilibrium Thermodynamics}, 
(North-Holland, Amsterdam, 1962).


\bibitem{Glansdorff71}
P. Glansdorff and I. Prigogine, 
{\sl Thermodynamics of Structure, Stability and Fluctuations}, 
(Wiley, New York, 1971).



\bibitem{Maes03}
Maes, C., Netocn\'y, K. 
{\sl Time-Reversal and Entropy}, 
Journal of Statistical Physics {\bf 110}, 269 (2003).

\bibitem{Maggi17}
U. M. B. Marconi, A. Puglisi, and C. Maggi, 
{\sl Heat, temperature and clausius inequality in a model for active Brownian particles},
Sci. Rep. {\bf 7}, 46496 (2017).



\bibitem{Marconi17}
 A. Puglisi and U. M. B. Marconi, 
 {\sl Clausius relation for active particles: What can we learn from fluctuations}, 
 Entropy {\bf 19}, 356 (2017).






\bibitem{Kubo66}
R. Kubo,
{\sl The fluctuation-dissipation theorem},
Rep. Prog. Phys. {\bf 29}, 255 (1966).




\bibitem{Felderhof78}
B U Felderhof, 
{\sl On the derivation of the fluctuation-dissipation theorem},
J. Phys. A: Math. Gen. {\bf 11}, 921 (1978).

\bibitem{Bhattacharjee}
J. K. Bhattacharjee, 
{\sl Elements of Nonequilibrium Statistical Mechanics}, 
Ane Books Pvt. Ltd, New Delhi, 2009.  















%\bibitem{Kardar15}
%A. P. Solon, Y. Fily, A. Baskaran, M. E. Cates, Y. Kafri, M. Kardar, and J. Tailleur, 
%{\sl Pressure is not a state function for generic active fluids}, 
%Nature Physics {\bf 11}, 673 (2015). 








%%%%  AOUP model  %%%
\bibitem{Szamel14}
G. Szamel, 
{\sl Self-propelled particle in an external potential: Existence of an effective temperature},
Phys, Rev. E {\bf 90}, 012111 (2014).

\bibitem{Cates21}
D. Martin, J. O'Byrne, M. E. Cates, \'E. Fodor, C. Nardini, J. Tailleur, and F. van Wijland
{\sl Statistical mechanics of active Ornstein-Uhlenbeck particles}, 
Phys. Rev. E {\bf 103}, 032607 (2021).

%%%% 
%\bibitem{Klymko17}
%D. Mandal, K. Klymko, and M. R. DeWeese, 
%{\sl Entropy Production and Fluctuation Theorems for Active Matter}, 
%Phys. Rev. Lett. {\bf 119}, 258001 (2017). 

%\bibitem{Lorenzo18}  
%Lorenzo Caprini, Umberto Marini Bettolo Marconi, Andrea Puglisi, and Angelo Vulpiani,
%{\sl Comment on "Entropy Production and Fluctuation Theorems for Active Matter"},
%Phys. Rev. Lett. {\bf 121}, 139801 (2018). 


\bibitem{Sevilla19}
F. J. Sevilla, R. F. Rodríguez, and J. R. Gomez-Solano, 
{\sl Generalized Ornstein-Uhlenbeck model for active motion}, 
Phys. Rev. E {\bf 100}, 032123 (2019).

\bibitem{Kaarakka11}
T. Kaarakka and P. Salminen, 
{\sl On fractional Ornstein-Uhlenbeck processes}, 
Communications on Stochastic Analysis. {\bf 5},  121 (2011).






\bibitem{Dhar20}
K. Malakar, A. Das, A. Kundu, K. V. Kumar, and A. Dhar, 
{\sl Steady state of an active Brownian particle in a two-dimensional harmonic trap}, 
Phys. Rev. E {\bf 101}, 022610 (2020).



\bibitem{Kramer}
H.A. Kramers,  
{\sl Brownian motion in a field of force and the diffusion model of chemical reactions}, 
Physica {\bf 7}, 284 (1940).



\bibitem{Pomeau17}
Y. Pomeau, J. Piasecki, 
{\sl The Langevin equation},
Comptes Rendus Physique, {\bf 18}, 570 (2017).




\bibitem{Dhar19}
A. Dhar, A. Kundu, S. N. Majumdar, S. Sabhapandit, and G. Schehr
{\sl Run-and-tumble particle in one-dimensional confining potentials: Steady-state, relaxation, and first-passage properties}, 
Phys. Rev. E {\bf 99}, 032132 (2019). 















%\bibitem{Frydel21a}
%D. Frydel, 
%{\sl Stationary distributions of propelled particles as a system with quenched disorder}, 
%Phys. Rev. E {\bf 103},  052603 (2021).

\bibitem{Basu20}
U. Basu, S.N. Majumdar, A. Rosso, S. Sabhapandit and G. Schehr, 
{\sl Exact stationary state of a run-and-tumble particle with three internal states in a harmonic trap}, 
J. Phys. A: Math. Theor. {\bf 53}, 09LT01 (2020).



















%\bibitem{Schnakenberg76}
%J. Schnakenberg, 
%{\sl Network theory of microscopic and macroscopic behavior of master equation systems},
%Rev. Mod. Phys. {\bf 48}, 571 (1976).
%
%\bibitem{Gaspard04a}
%P. Gaspard, 
%{\sl Fluctuation theorem for nonequilibrium reactions}, 
%J. Chem. Phys. {\bf 120}, 8898 (2004).
%
%\bibitem{Gaspard04b}
%P. Gaspard, 
%{\sl Time-Reversed Dynamical Entropy and Irreversibility in Markovian Random Processes}, 
%J. Stat. Phys. {\bf 117}, 599 (2004).
%
%\bibitem{Andrae10}
%B. Andrae, J. Cremer, T. Reichenbach, and E. Frey, 
%{\sl Entropy Production of Cyclic Population Dynamics}, 
%Phys. Rev. Lett. {\bf 104}, 218102 (2010).
%
%\bibitem{Lan12}
%G. Lan, P. Sartori, S. Neumann, V. Sourjik, and Y. Tu, 
%{\sl The energy–speed–accuracy trade-off in sensory adaptation}, 
%Nat. Phys. {\bf 8}, 422 (2012).
%
%\bibitem{Li19}
%J. Li, J. M. Horowitz, T. R. Gingrich, and N. Fakhri, 
%{\sl Quantifying dissipation using fluctuating currents}, 
%Nat.Commun. {\bf 10}, 1666 (2019).

%D. M. Busiello and A. Maritan, J. Stat. Mech.: Theory Exp.
%(2019) 104013.






%%%  RTP in 1D model  %%%
\bibitem{Schnitzer93}
 M. J. Schnitzer, {\sl Theory of continuum random walks and application to chemotaxis}, Phys. Rev. E {\bf 48}, 2553 (1993).

\bibitem{Angelani17}
L. Angelani, {\sl Confined run-and-tumble swimmers in one dimension}, 
J. Phys. A: Math. Theor. {\bf 50}, 325601 (2017).

\bibitem{Dhar18}
K. Malakar, V. Jemseena, A. Kundu, K. V. Kumar, S. Sabhapandit, S. N. Majumdar, S. Redner, and A. Dhar, 
{\sl Steady state, relaxation and first-passage properties of a run-and-tumble particle in one-dimension}, 
J. Stat. Mech.: Theory Exp. 043215 (2018).

\bibitem{Frydel22b}
D. Frydel,
{\sl The four-state RTP model: exact solution at zero temperature}, 
Phys. Fluids {\bf 34}, 027111 (2022).



\bibitem{rudi18}
D. Frydel and R. Podgornik, {\sl Mean-field theory of active electrolytes: Dynamic adsorption and overscreening}, 
Phys. Rev. E {\bf 97}, 052609 (2018).























\end{thebibliography}
\end{document}